**Classification**: *Biological Sciences; Biophysics and Computational Biology, Evolution*

# Soluble oligomerization provides a beneficial fitness effect on destabilizing mutations.

Shimon Bershtein[a], Wanmeng Mu[a,b], and Eugene I. Shakhnovich[a1]

[a]*Department of Chemistry and Chemical Biology, Harvard University, 12 Oxford St, Cambridge, MA 02138*
[b]*State Key Laboratory of Food Science and Technology, Jiangnan University, Wuxi 214122 P. R. China*

## Abstract

Mutations create the genetic diversity on which selective pressures can act, yet also create structural instability in proteins. How, then, is it possible for organisms to ameliorate mutation-induced perturbations of protein stability while maintaining biological fitness and gaining a selective advantage? Here we used a new technique of site-specific chromosomal mutagenesis to introduce a selected set of mostly destabilizing mutations into *folA* - an essential chromosomal gene of E. *coli* encoding dihydrofolate reductase (DHFR) - to determine how changes in protein stability, activity and abundance affect fitness. In total, 27 E.coli strains carrying mutant DHFR were created. We found no significant correlation between protein stability and its catalytic activity nor between catalytic activity and fitness in a limited range of variation of catalytic activity observed in mutants. The stability of these mutants is strongly correlated with their intracellular abundance; suggesting that protein homeostatic machinery plays an active role in maintaining intracellular concentrations of proteins. Fitness also shows a significant correlation with intracellular abundance of soluble DHFR in cells growing at 30$^{\rm o}$C. At 42$^{\rm o}$C, on the other hand, the picture was mixed, yet remarkable: a few strains carrying mutant DHFR proteins aggregated rendering them nonviable, but, intriguingly, the majority exhibited fitness *higher* than wild type. We found that mutational destabilization of DHFR proteins in E. *coli* is counterbalanced at 42$^{\rm o}$C by their soluble oligomerization, thereby restoring structural stability and protecting against aggregation.

\body
**Introduction**

In order to evolve, an organism must acquire genetic mutation(s), yet these very mutations can cause structural destabilization of the proteins they encode(1-4), potentially affecting the ability of an organism to survive and reproduce (fitness). This dichotomy of how cells can accommodate evolutionarily beneficial, but structurally destabilizing mutations (the 'genotype-phenotype gap') is central to Biology, yet poorly understood.

Past efforts to bridge the genotype-phenotype gap 'in one shot'' – from sequences to phenotype(5) may have fallen foul of the indirect relationship between genomic sequences and phenotypic traits, where many sequence variations lead to the same phenotypic effect. However, the relationship between genomic sequences and phenotypic traits is likely to be indirect with many sequence variations leading to the same phenotypic effect. A more promising approach, introduced in recent multi-scale theoretical models (6-8) is to bridge the scales gap mid-way by relating coarse-grained molecular traits to organism's fitness.

Of all molecular traits, protein stability has been widely recognized as one of the most evolutionarily important (3, 7, 9-11). Indeed, to be functional, almost all proteins must be either stably folded or, in the case of many intrinsically-disordered proteins, assume a specific structure upon binding to a partner (12). However, the relationship between stability and fitness, while often postulated (9, 10) (11), has not been sufficiently explored experimentally.

As such we chose to partially close the phenotype-genotype gap by exploring the relationship between stability of an essential enzyme, dihydrofolate reductase (DHFR) encoded by the *folA* gene in *Escherichia coli*, and bacterial fitness. DHFR catalyzes an electron transfer reaction from NADPH to 7,8-dihydrofolate ($H_2F$) to form 5,6,7,8-tetrahydrofolate ($H_4F$). DHFR struck us as a good choice for this investigation since tetrahydrofolate is essential for the synthesis of purines, thymidylate, and several amino acids, (13), and DHFR is present in relatively low abundance in *E.coli* (~40 copies/cell (14)), so toxicity from its aggregation should be negligible. It was reasonable therefore to expect that the main effect of perturbations in fitness would be mediated *via* modulation of the active enzyme copy number.

With this in mind, we explored the effects of mutations in DHFR that, theoretically, would have minimal affects on enzyme activity, yet provide a broad range of affects on the fitness of our chosen *E.coli* host organism. Unlike earlier approaches, which considered how over-expression of non-endogenous destabilized proteins from a ***plasmid*** affects fitness (5, 15, 16), we aimed to establish the link between the molecular properties of an *endogenous* essential protein and organismal *fitness* at maximally realistic conditions through site-specific mutations carried out directly on the chromosome while leaving intact the regulatory region responsible for control of endogenous expression levels (Fig. 1). Through this design, we hoped to determine which particular molecular properties of DHFR, have the most pronounced impact on the fitness of *E. coli*.

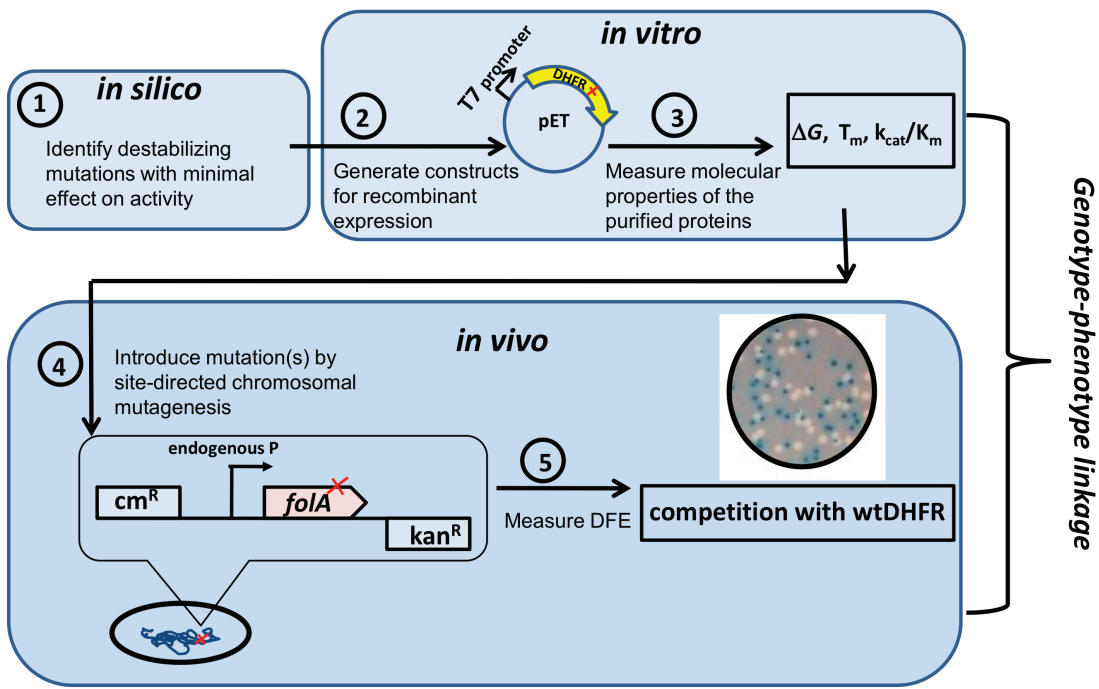

**Figure 1. The experimental approach**. (1) 10 DHFR residues, predominantly from the hydrophobic core and at least 4Å from the NADPH and dihydrofolate binding sites, were chosen for mutagenesis based on the structural and phylogenetic predictions and published biophysical and biochemical data. (2) 16 single mutants (Table 1) were generated, cloned into pET vector, expressed and purified. (3) Gibbs free energy difference between folded and unfolded states ($\Delta G$), apparent mid-transition temperature of unfolding ($T_m^{app}$), and catalytic parameters ($k_{cat}$, $K_m$) were measured (Table 1). (4) A site-directed chromosomal mutagenesis method was developed to introduce in vitro characterized mutations into chromosomal folA gene of E. coli's MG1655 strain without perturbing the gene's regulatory region. In addition to 16 single mutant strains, 11 multiple mutant strains were generated by combining exhaustively 4 most destabilizing mutations (V75H, I91L, W133F, I155A) (Table S1). (5) Fitness effects of the introduced mutations (in total 27 strains) were measured by growth competition of mutant strains with wtDHFR strain.

## Results

*Selection and in vitro analysis of DHFR mutants.* The rationale behind choosing DHFR mutations was to cover a broad range of stabilities with minimum effect on activity. To that end, an extensive search of the literature identified 10 loci, all of which were distant (at least 4Å away) from the enzymatically critical NADPH and $H_2F$ binding sites - most being buried in the protein's hydrophobic core (Table 1). Multiple sequence alignments at the selected loci were examined and substitutions with both low and high conservation propensity were identified (Table 1, Fig. S1).

**Table 1. Thermodynamic, structural, and catalytic properties of the *in vitro* purified DHFR mutants.**

| Mutant | $T_m^{app}$ (°C) | $\Delta G_{(H2O)}$ (kcal/mol) 25°C | $k_{cat}$ (s$^{-1}$) | $k_{cat}/K_m$ (s$^{-1}$µM$^{-1}$) | ASA[a] | Conservation[b] (%) native | substitution |
|---|---|---|---|---|---|---|---|
| wt | 51.7 | -4.4 | 11.65 | 3.6 | | | |
| V40A | 43.2 | -3.31 | 24.05 | 3.22 | 0 | 43.6 | 0.34 |
| I61V | 53.4 | -4.52 | 8.99 | 3.7 | 0 | 19.9 | 79 |
| V75H | 40.6 | -2.18 | 15.5 | 4.57 | 0.05 | 32 | 0 |
| V75I | 39.2 | -2.98 | 18.25 | 5.51 | 0.05 | 32 | 10.3 |
| I91V | 49.1 | -3.14 | 17.87 | 5.13 | 0.07 | 33.3 | 29 |
| I91L | 41.4 | -2.69 | ND[c] | ND[c] | 0.07 | 33.3 | 20 |
| L112V | 47.4 | -2.7 | 11.45 | 2.34 | 0 | 34 | 27.5 |
| W133F | 46.3 | -4.4 | 13.45 | 6.79 | 0.05 | 53.6 | 20.6 |
| W133V | ND[c] | -1.53 | ND[c] | ND[c] | 0.05 | 53.6 | 0.34 |
| I155T | 43.4 | -3.68 | 11.27 | 2.72 | 0.12 | 15.8 | 21 |
| I155L | 45.8 | -2.84 | 12.9 | 5.12 | 0.12 | 15.8 | 1 |
| I155A | 38.5 | -2.42 | 11.95 | 2.71 | 0.12 | 15.8 | 1 |
| I115V | 51.4 | -6 | 10.26 | 2.08 | 0.02 | 55.3 | 34.4 |
| I115A | 46.1 | -3.4 | 7.59 | 0.5 | 0.02 | 55.3 | 0 |
| V88I | 44.3 | -4.22 | 13.8 | 3.12 | 0.34 | 12.4 | 1.4 |
| A145T | 51.3 | -4.13 | 10 | 3.6 | 0.89 | 12.4 | 1.4 |

[a]ASA (accessible surface area) was calculated by Vadar package (http://vadar.wishartlab.com) ; cutoff for buried residues is around 0.25
[b]Conservation (%) of a native E. coli's DHFR residue (left column) or a substituted residue (right column) in a given position of 290 aligned mesophylic prokaryotic DHFR sequences retrieved from the Optimal Growth Temperature database (http://pgtdb.csie.ncu.edu.tw)
[c]Not determined

Using site-directed mutagenesis we generated 16 constructs for recombinant expression of DHFR, each carrying a single unique mutation in the coding region. All DHFR mutants were expressed and purified and their biophysical and catalytic properties assayed initially as follows: (i) Thermal stabilities of the mutants were measured by Differential Scanning Calorimetry (DSC) and the obtained thermograms used to infer apparent thermal transition midpoint temperatures ($T_m^{app}$) (Table 1, Fig. S2). (ii) A two-state folding model was applied to urea denaturation curves to derive urea mid-transition concentration ($C_m$), and the Gibbs free energy difference between folded and unfolded states at 25°C in water ($\Delta G_{H2O}$) (Table 1, Fig. S3, Table S1).

Both $C_m$ and $\Delta G_{H2O}$ were linearly correlated to $T_m^{app}$ in the studied range of temperatures (30-42°C), as postulated by protein thermodynamics (17, 18) (Fig. S4). As expected, most mutations (13 out of 16) appeared to be mildly to severely destabilizing with apparent $\Delta\Delta G$ values ranging between 0.18 to 2.87 kcal/mol (Table S1). This initial work led us to choose to use $T_m^{app}$ to characterize the stability of mutants, as it could be directly measured experimentally, in contrast to $\Delta G_{H2O}$ which is derived under the additional assumption of two-state unfolding.

In terms of catalytic activity, the $k_{cat}$ and $K_M$ of the DHFR proteins were measured by full progress-curve kinetics (Table 1). As expected, the enzymatic proficiency ($k_{cat}/K_M$) of most mutants was found to be very similar to that of the wt DHFR (with only a 2 fold difference between all mutants). There was no statistically significant correlation between protein activity and stability (Fig. S4), suggesting a lack of trade-off for this group of mutations. Further analysis also showed no significant correlation between catalytic activity and fitness (Fig. S4), suggesting a catalytic saturation regime for DHFR (as found by Hartl and co-workers for another enzyme(19)).

In addition to 16 single DHFR mutants, we built 11 constructs carrying multiple mutations. This was achieved by exhaustively combining the 4 most destabilizing single mutations V75H, I91L, W133V, and I155A (Table 1, Table S2). However, we were unable to purify these mutants at quantities required for *in vitro* characterization.

*Site-directed chromosomal mutagenesis.* Chromosomal mutants were created by a technique developed in this lab that allows incorporation of desired mutations in a controlled manner at any locus within *E. coli's* chromosome (see Materials and Methods).

Using this technique, we generated 27 *folA* mutants in *E. coli*'s MG1655 strain (16 single mutants with defined molecular properties (Table 1) and 11 multiple mutants (Table S2)).

*Fitness measurements.* Fitness of these strains was determined by competition with wtDHFR strain following the assay developed by Lenski and co-workers (20). To this end, wt and a given mutant strain were mixed in 1:1 ratio and grown together for 18 hours in a range of temperatures. Laboratory fitness of a mutant strain in this assay was defined as the population ratio of mutant strain to wt upon completion of the growth cycle (see Materials and Methods). Below we focus on 3 temperatures; 30°C as the lower limit of Arrhenius-like dependence of growth rate on temperature (**Fig. S5**), 42°C as its highest limit, and 37°C as a '' physiological'' temperature for *E. coli* in humans. We viewed this competition assay as superior to other fitness measurements (such as individually measured growth rates), as it allows the determination of an evolutionarily relevant selective (dis)advantage of a population competing for nutrients. Two factors - *experimental noise* and *fitness variations* due to the genetic diversity *outside* of the *folA* locus in competing strains - can limit the accuracy of fitness measurements from competition assays. Two steps were therefore taken to establish the error bars for fitness measurements arising from these issues.

First, *experimental noise* was measured for any competing pair of strains by repeating the competition experiments 2-6 times. 'Noise' levels were found to be within 13% for measurements performed at 30°C, and 17% at 42°C (Fig. 2-3). Second, although wtDHFR strains, just as the wtMG1655 strain, do not carry mutation in the *folA* gene, they were subjected to double antibiotic selection, and, therefore, could carry unrelated hitchhiked mutations in other parts of the genome; (see Material and Methods). It is possible, therefore, that '*fitness variations*' could arise due to genetic variation *outside* of the *folA* locus in competing strains. This was addressed by using our chromosomal insertion technique to generate an additional twelve control ''wt'' strains whose *folA* gene encodes wtDHFR to measure any 'unintended' genetic variability that this technique might introduce elsewhere in the genome. 12 individual competition assays were carried out by competing each of the new control wtDHFR strains with our original wtDHFR reference strain. We found that the experimental error for these competitions was comparable to the error found for any repeatedly competed pair of strains (10% for competitions at 30°C, and 18% at 42°C) (Table S3). This analysis establishes the error bars for fitness measurements.

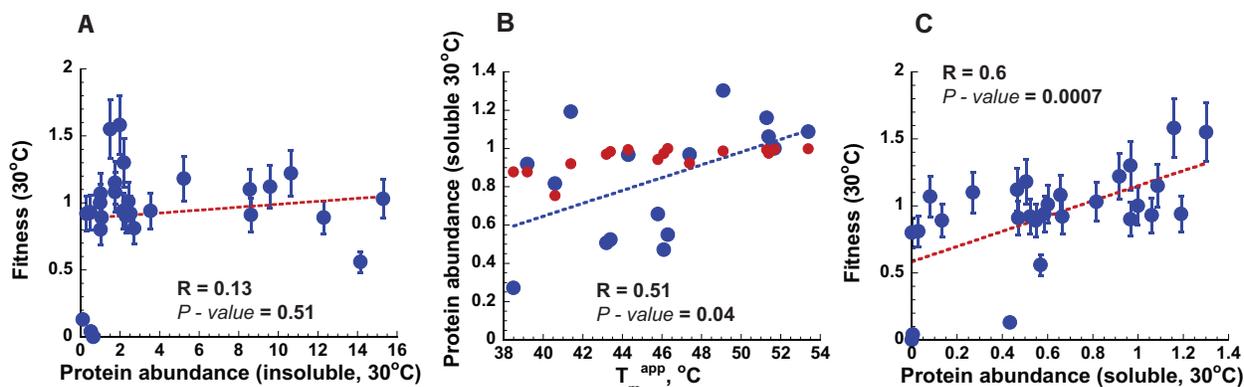

**Figure 2. Correlating fitness at 30°C with molecular properties**. **A**, Fitness measured at 30°C by competing 27 DHFR mutant strains with a reference wtDHFR strain is plotted against insoluble fraction of DHFR proteins determined in cell lysates by Western blot. Fitness of the wtDHFR strain and its abundance in the insoluble fraction is set to 1. **B,** Soluble protein abundance (blue circles) determined for single DHFR mutant strains is plotted against apparent mid-transition temperature of unfolding ($T_m^{app}$) measured in vitro by DSC. Protein abundance predicted at 30°C by Equation (1) is depicted in red. Predicted fraction somewhat deviates from a plateau because of the experimental imprecision of ΔG values derived from the urea unfolding under two-state assumption and $T_m^{app}$ values inferred from the DSC thermograms (Fig. S4). **C**, Fitness measured at 30°C by competing 27 DHFR mutant strains with a reference wtDHFR strain is plotted against soluble fraction of DHFR proteins determined in cell lysates by Western blot. Fitness of the wtDHFR strain and its abundance in the soluble fraction is set to 1.

*Fitness correlates with protein abundance at 30°C.* We measured intracellular abundances of DHFR proteins for all mutants in both soluble and insoluble fractions of cell lysates. There was no significant correlation between fitness and abundance in the *insoluble* fraction at all temperatures tested (30°C, 37°C, and 42°C), confirming the notion that for this protein toxicity of the aggregated fraction does not contribute to fitness (Fig. 2A, Fig S6). *Soluble* protein abundance however does appear to be correlated to stability. Correlation of the soluble fraction of the DHFR single mutant proteins with their $T_m^{app}$ grows significantly stronger as temperature increases: R = 0.51 (30°C), 0.64 (37°C), and 0.76 (42°C) (Fig. 2B, Fig. S6). Statistical mechanics of two-state protein folding (21, 22) would predict the relationship between protein stability and intracellular abundance of folded proteins to follow the Boltzmann law:

$$C_f = \frac{C_{tot} e^{-\frac{\Delta G}{k_B T}}}{\left(1 + e^{-\frac{\Delta G}{k_B T}}\right)} \quad (1)$$

Where T is temperature, $k_B$ is Boltzmann constant, $C_{tot}$ and $C_f$ are total and folded abundances. The Boltzmann relationship in Eq (1) predicts a plateau for the observed range of free energies of mutants at $\Delta G \leq -2 kcal/mol$ at 30°C (Table 1), provided that total abundance is fixed (red circles, Fig. 2B, Fig. S6). Our data, however, show a stronger, linear, dependence between abundance and stability (blue circles, Fig. 2B, Fig. S6), suggesting that cytoplasm is an active medium, where stability critically affects total abundance through homeostatic balance between protein production and degradation. Destabilization of a protein apparently shifts this balance towards degradation - as has been observed *in vitro* (23, 24). In line with this observation is the strong correlation between fitness and protein abundance observed at 30°C (Fig. 3B). However, we found that this correlation disappears at higher temperature (Fig. S6).

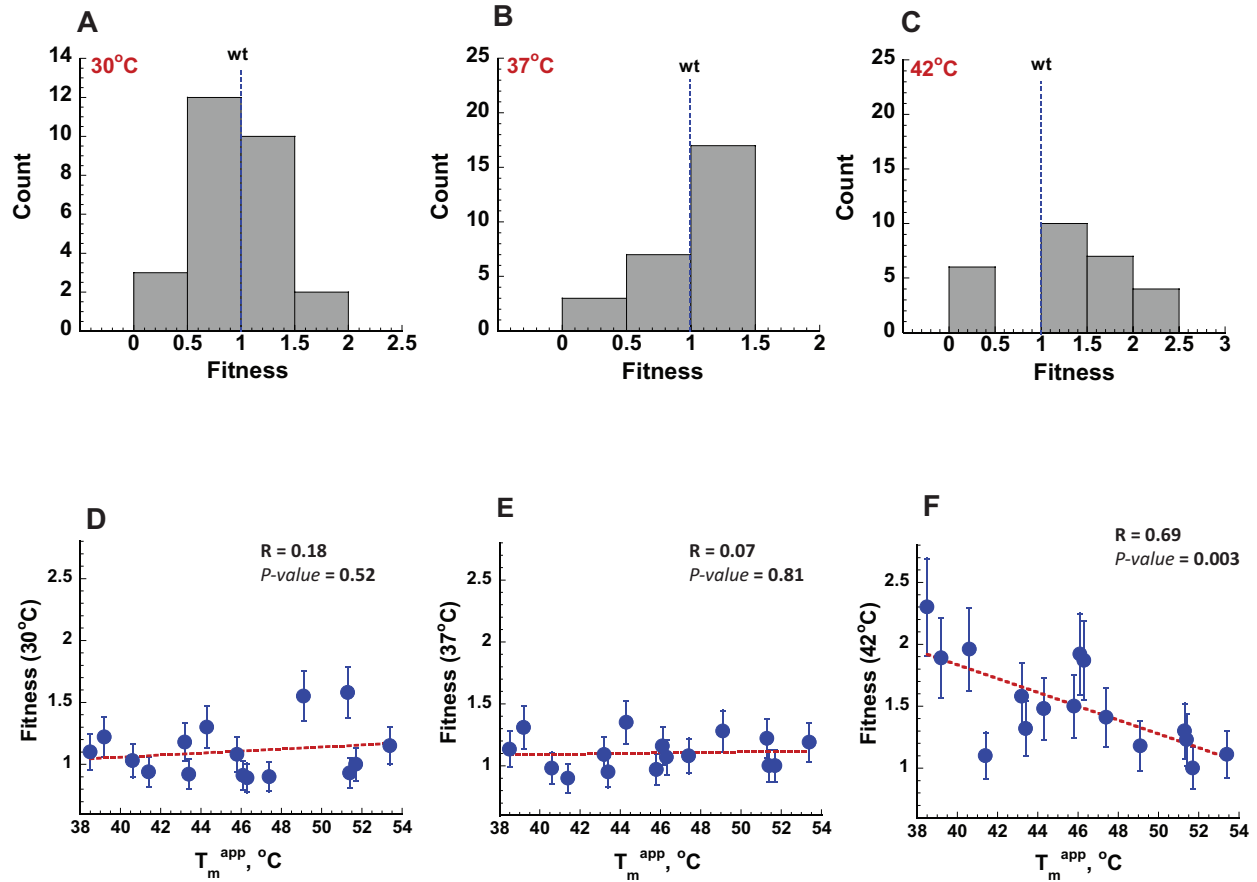

**Figure 3. Distribution of fitness effects and correlation with stability. A, B, C**, Histograms of distributions of fitness effects at 30°C, 37°C, and 42°C for all mutant DHFR strains (16 single

and 11 multiple mutants) as measured by competition with a reference wtDHFR strain. Fitness of the reference strain is set to 1. Individual fitness values can be found at Table S2. **D, E, F**, Correlation of fitness values measured for single mutant DHFR strains by competition at 30°C, 37°C, and 42°C to *in vitro* measured $T_m^{app}$.

*Fitness inversely correlates with stability at 42°C.* At 30°C, most (over 55%) of the DHFR mutations were deleterious or nearly-neutral (Fig. 3A, Table S2). Very surprisingly, with temperature *increase* the fraction of strains carrying deleterious mutations *dropped* to 27% at 37°C, and to only 14% at 42°C (Fig. 3B-C, Table S2). Moreover, the Distribution of Fitness Effects (DFE) became practically bi-modal at 42°C: The effect of DHFR mutations was either nearly-lethal or, for most strains, clearly advantageous. We confirmed this unusual observation by competing I155A DHFR mutant strain (that shows highest fitness level at 42°C among all DHFR mutants, Table S2) with 9 independently generated wtDHFR strains. In line with our previous observation, at a low temperature (30°C), the I155A mutant exhibited reduced fitness (0.92), while at 42°C its fitness improved remarkably, on average 2 fold (1.89) compared to wt (Table S4). While at 30°C and 37°C no statistically significant correlation between $T_m^{app}$ and fitness can be found, there does appear to be a clear *anti-correlation* at 42°C (Fig. 3D-F), implying an unusual trade-off between fitness and stability - *i.e.* mutant strains encoding *less* stable DHFR appear to be *more* fit at the higher temperature (Fig. 3D-F).

*Soluble oligomerization at 42°C.* The results of our fitness experiments presented a paradoxical situation where the main determinants of fitness reverse themselves for strains growing at higher temperature. While the observed dependence of fitness on protein abundance and activity at 30°C is intuitive and consistent with theoretical views based on flux balance theory (19, 25), the data at 42°C where strains carrying less stable mutants appear to be more fit are totally unexpected (and, on the face of it, counterintuitive). What, then, is going on at 42°C that favours strains carring less stable mutants? A critical hint comes from the hyperthermophile *Thermotoga Maritima* whose DHFR exists as a stable homodimer (26). Further, Fernandez and Lynch recently observed that less stable proteomes are enriched in protein complexes (27). This work suggests a possibility that, at higher temperatures, destabilized mutants can form soluble oligomers preventing their further aggregation and preserving activity. Indeed, such behavior at elevated temperatures close to unfolding transition has been predicted for another protein, SH3 domain (28) -furthermore, simulations for several proteins (28-30) and experimental inverstigations (31) have indicated that oligomerization can increase stability by providing additional stabilizing contacts, especially in domain-swapped oligomers. With this in mind, we sought to evaluate, at the protein level, the propensity of DHFR mutants to oligomerize *in vitro*; initially by solution FRET, then by *in vitro* cross-linking experiments at 'high' (42°C) and 'low' (25°C) growth temperatures.

Solution FRET was carried out with Cy3 as a donor label and Cy5 as an acceptor to determine whether mutant DHFR associates *in vitro* at 42°C. We found that the highly destabilized mutant I155A, which gives rise to the most fit (at 42°C) strain, shows pronounced protein-protein interaction at 42°C in contrast to wt (also at 42°C) and the same mutant at 25°C – matching our observations on fitness (Fig. S7). However, FRET data for another destabilized mutant, W133V, whose strain has very low fitness at 42°C was qualitatively similar to I155A and different from wt. (Fig. S7). This latter anomaly was thought to be due to a limitation of FRET in its inability to distinguish between oligomers and aggregates as FRET merely monitors a close proximity between donor and acceptor dyes attached to different protein molecules. We hypothesized that two distinctly different types of fitness effects as seen in Fig. 3C at 42°C may be due to the differences in the association behavior of different mutants (an effect FRET is incapable of identifying). To test this hypothesis, we carried out *in vitro* cross-linking experiments at room temperature (25°C) and at 42°C on the purified DHFR proteins (Fig. 4, Fig. S8). Strikingly, we found that many mutants that exhibit higher fitness at 42°C (e.g. I155A) have a strong tendency to form oligomers at 42°C but not at room temperature (Figure 4A). Furthermore, oligomer-forming mutants did not exhibit high molecular weight bands typical of aggregated proteins, while mutants of low fitness (at 42°C) strains showed a pronounced aggregation and a much weaker homodimer band (Fig. 4, Fig. S8). Quantitatively, we found a significant correlation between the propensity to oligomerize (assessed by density of all DHFR oligomerized species bands in the gels) with both fitness at 42°C (Fig. 4B) and unfolding transition temperature $T_m$ (Fig. 4C). Further to support our postulation of soluble oligomerization is provided by native gel electrophoresis (Fig. S9), where the destabilized I155A mutant shows same pattern of dimers at 42°C but not at lower temperatures, a behavior not observed for wt proteins.

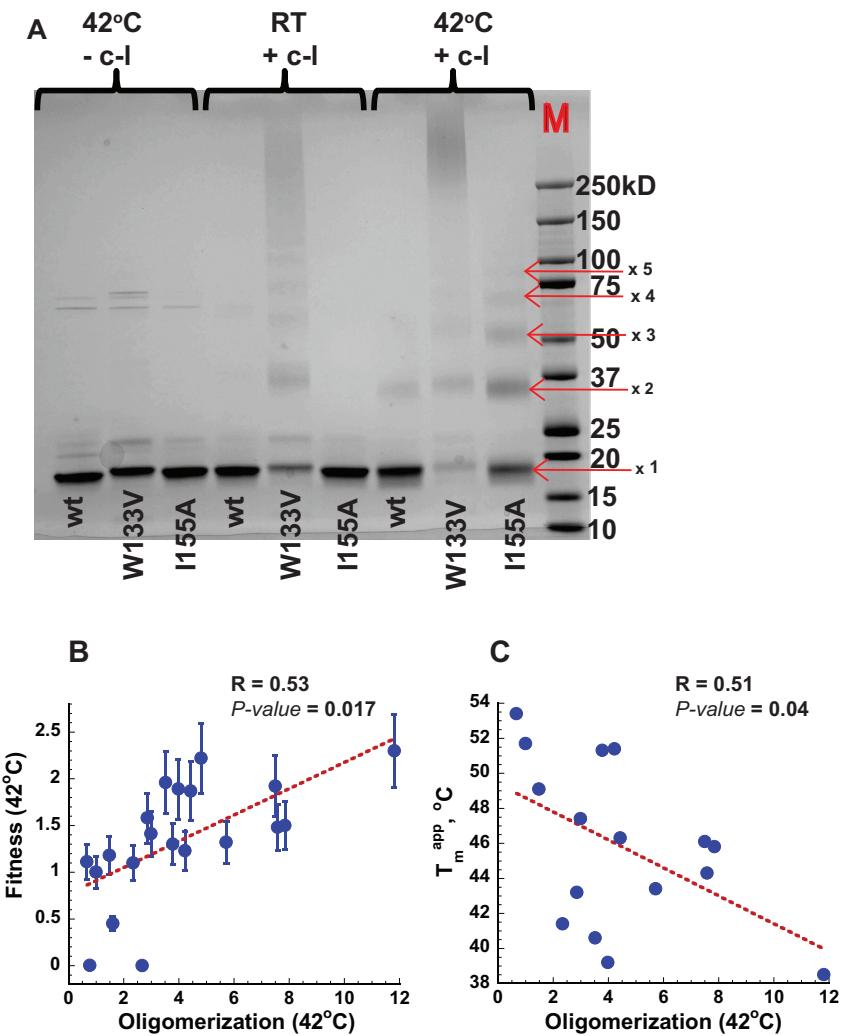

**Figure 4. *In vitro* oligomerization assay**. **A**, A representative cross-linking experiment. 11 µM of wt DHFR, W133V, or I155A purified proteins in 25 mM potassium-phosphate buffer (pH 7.8) and 100 µM NADPH were incubated for 45 min at room temperature (RT) or 42°C with or without 2 mM of cross-linking (c-l) agent (glutaraldehyde). Shown is the SDS-PAGE analysis of 10 µl of protein samples after Coomassie staining. Red arrows indicate molecular weight equivalent of monomeric and oligomeric protein species. Smears at high molecular weight seen for W133V protein at 37°C and 42°C in the presence of c-l are due to extensive aggregation. **B**, The *in*

*vitro* propensity to oligomerize is correlated to fitness values measured by competition at 42°C. Oligomerization propensity was determined by measuring the relative density of the SDS-PAGE bands corresponding to *all* DHFR oligomerized species for each of the purified mutant (Fig. S8). The degree of oligomerization for every mutant is normalized to wt DHFR. **C**, Correlation of oligomerization propensity (as measured in **B**) to mutant's $T_m^{app}$.

**Discussion**

A key aspect of this study is that it introduces mutations in the gene of interest directly on the chromosome leaving the upstream region intact to keep protein abundance at its endogenous level, including possible feedback control mechanisms. This contrasts with the more common approach where DHFR mutant genes are expressed from a plasmid. This is problematic as effects observed in this work are strongly concentration-dependent so that non-endogenous abundance of mutant proteins would have concealed or distorted the relevant phenomenology.

A common expectation is that destabilization should be detrimental to fitness, firstly, because active proteins are thought to be lost to unfolding (see Eq(1)) and, secondly, because destabilized proteins are believed to aggregate causing both irreversible loss of function and possible toxicity (8, 32). Our results suggest that loss of function rather than toxicity from aggregation is responsible for most fitness effects at 30°C (Fig. 2). While this finding might appear at variance with recent theoretical postulates that posit that misfolding-induced aggregation is a major source of fitness effects (8), we note that DHFR is a low copy number enzyme. As such it is possible that the effect of misfolding-induced aggregation may be more pronounced for highly expressed proteins as it was indeed observed in recent experiments where a non-endogenous protein was greatly overexpressed from a plasmid in Yeast (16). However, the behavior, which we report at a higher temperature (42°C), goes against the conventional wisdom because an unexpected physical factor interferes. Our works shows that less stable DHFR proteins tend to escape aggregation by forming soluble oligomers, and this phenomenon is responsible for a surprising anti-correlation between stability and fitness at 42°C.

Dimer stabilization through domain swapping has been observed *in silico* and *in vitro* for many proteins (28, 33, 34). Recent observations that homodimers are prevalent in proteomes (35) may indicate that soluble oligomerization serves as a common evolutionary mechanism to evolve stable functional proteins (36, 37). It appears that this mechanism may provide a route to escape the detrimental effects of destabilizing mutations by opening sequence space for a broader exploration at higher temperatures. It is tempting to suggest that it can serve as a universal route to ''sequence-based'' (38) thermal adaptation for many proteins of originally mesophilic organisms which colonize warmer environments. While the structural aspects of mutant DHFR dimerization remain to be discovered, our findings provide a mechanistic support to this view.

We found here a peculiar temperature-dependent tradeoff between stability and fitness, which is very different from the postulated tradeoff between stability and activity (1). For example, the destabilized mutant I155A is less fit than wt at 30°C when it cannot dimerize via domain swapping, and the effect of destabilization is in a decreased copy number of active proteins. However, the same mutant is much more fit than wt at 42°C where soluble oligomerization of the same I155A mutant prevents its aggregation while wt DHFR partly aggregates.

This study highlights the complexity of the concept of fitness, which is central to population genetics. Mutations that provide higher fitness under one set of conditions can be detrimental under another. It seems that the fate of a mutation is determined not by its fitness effect under fixed conditions but rather by an organism's lifestyle (e.g. generalist vs. specialist). Indeed some

DHFR mutations, which provide high fitness to MG1655 *E.coli* at $42^0$C are wild-type in other bacterial species.

The metaphor of a 'rugged fitness landscape' is often invoked to reflect the notion that fitness effects cannot be predicted from sequence variation. This study shows that fitness effects of mutations, while difficult to rationalize at the level of sequence variation, can be ''projected'' on a small number of ''axes'' that reflect coarse-grained Biophysical properties of proteins such as stability or intracellular abundance. Organismal fitness in the space of such coarse-grained properties can then appear more ''smooth''. Indeed, a significant correlation exists between fitness and several coarse-grained quantities. This is good news for Biophysics-based multiscale modeling of evolution. The bad news is that it is still challenging to postulate such correlations a'priori – for example, simple equilibrium Statistical Mechanics considerations may be not applicable because cellular environments represent an active medium where energy consuming machinery acts on proteins. Further issues may be encountered where unexpected equilibrium mechanisms, such as soluble oligomerization may intervene in organismal fitness.

**Materials and Methods**

*Protein expression and purification*

*folA* genes carrying the mutations and C-terminal His-tag were cloned in a pET24 expression vector under the inducible T7 promoter, then transformed into BL21(DE3) cells and expressed by a standard protocol. The recombinant proteins were purified on Ni-NTA columns (Qiagen) to over 95% purity.

*Enzyme kinetics*

DHFR kinetic parameters were measured by progress-curve kinetics, essentially as described (39). Purified enzymes (10 nM) were pre-incubated with 120 µM NADPH in MTEN buffer (50 mM 2-(N-morpholino)ethanesulfonic acid, 25 mM tris(hydroxymethyl)aminomethane, 25mM ethanolamine, and 100 mM sodium chloride, pH7). The reaction was initiated by addition of dihydropholate (20, 15, 10 µM final concentration) and monitored by coenzyme fluorescence (ex. 290 nm, em. 450 nm) until completion (assays were performed at $25^oC$). The kinetics parameters ($k_{cat}$ and $K_M$) were derived from progress-curves analysis using Global Kinetic explorer(40).

*Stability measurements*

Thermal stability was characterized by Differential Scanning Calorimetry (DSC), essentially as described in (41). Briefly, DHFR proteins in Buffer A (10 mM potassium-phosphate buffer pH8.2 supplemented with 0.2 mM EDTA and 1 mM beta-mercaptoethanol) were subjected to a temperature increase of $1^oC$/min between 20 to $80^oC$ (nano-DSC, TA instruments), and the evolution of heat was recorded as a differential power between reference (buffer A) and sample (120 µM protein in buffer A) cells. The resulting thermogram (after buffer subtraction) was used to derive apparent thermal transition midpoints ($T_m^{app}$). Thermal unfolding appeared irreversible for all DHFR proteins tested, and was characterized by a complex transition that cannot be fitted to a simple two-state model (41). Urea unfolding was used to measure chemical stability of the DHFR mutants. Proteins (25 µM in buffer A) were diluted in urea (0.2 mM increments up to a final urea concentration between 0 and 6 M), pre-equilibrated over night at room temperature, and the change in the folded fraction was monitored by a circular dichroism signal at far-uv wavelength (221 nm) at $25^oC$(J-710 spectropolarimeter, Jasco). Fitting to a two-state model was used to derive the following: 1) chemical transition midpoint ($C_m$), 2) denaturant dependence (*m*-value), and 3) Gibbs free energy difference between folded and unfolded states in water ($\Delta G_{H2O}$) (42).

*Site-directed chromosomal mutagenesis*
The method is a modification of a chromosomal gene knock-out protocol (43). Briefly, the *folA* gene carrying the desired mutation(s) with an entire endogenous regulatory region (191 bp separating the stop codon of the upstream *kefC* gene and start codon of the *folA* gene) was placed on a pKD13 plasmid flanked by two different antibiotic markers (genes encoding kanamycin (kanR) and chloramphenicol (cmR) resistances). The entire cassette was then amplified with two primers tailed with 50 nucleotides homologous to the region of the chromosome intended for recombination (*kefC* gene upstream and *apaH* gene downstream to *folA)*. The amplified product was transformed into BW25113 strain with an induced Red helper plasmid. The recombinants were selected on plates carrying both antibiotics. Strains carrying the desired mutation in the chromosome were verified by sequencing. Identified chromosomal mutations were then re-transformed into MG1655 strain by P1 transduction and double antibiotic selection (kan and cm) and again verified by sequencing.

*Competition assay*
Laboratory fitness was determined by competing each of the strains expressing mutant DHFR protein with wt DHFR strain in M9 minimal media supplemented with 0.2% glucose, 1mM MgSO4, 0.1% casamino acids, and 0.5 μg/ml thiamine. To this end, MG1655 E. *coli* strain carrying wt *folA* gene flanked by cmR and kanR genes was mixed with one of the MG1655 DHFR mutant strains in a 1:1 ratio ($\approx 10^4$ cells each) in 50 ml of medium. Prior to mixing, cells were grown separately overnight from a single colony, diluted 1/100 and re-grown to early exponential phase ($OD_{600} \approx 0.2$) at $30^{\circ}C$. The competition assay was performed for 18 hours at $30^{\circ}C$, $37^{\circ}C$, and $42^{\circ}C$. To distinguish between the competing strains, a knock-out mutation was introduced in the *lacZ* gene of each strain. Two identical competition experiments were performed with the *lacZ* knockout (a neutral marker under the condition of the competition experiment) always present in one of the competing strains. The ratio before and after competition was determined by plating the culture on LB agar plates supplemented with X-gal and IPTG (*lacZ*⁻ strain generates white colonies, whereas *lacZ*⁺ strain generates blue colonies, hence "blue-white" swap). The swap was performed to ensure the neutrality of the *lacZ* marker. Around 3,000-5,000 colonies were counted for each competition experiment.

*Intracellular protein abundance*
Cells were grown in supplemented M9 medium for 6 hours at $30^{\circ}C$, chilled on iced for 30 min and lysed with BugBuster (Novagen). The insoluble fraction was separated by centrifugation, then solubilized by Inclusion Body Solubilization Reagent (Thermo Scientific), and dialyzed against PBS. DHFR amounts in the soluble and insoluble fractions were determined by SDS-PAGE followed by Western Blot using Rabbit-anti E.*coli*'s DHFR polyclonal antibodies (custom raised by Pacific Immunology) by measuring densities of the DHFR bands. All protein abundances are normalized to the amounts detected for wt DHFR then arbitrarily set to 1. The overall amount of protein in the lysates was estimated with BCA protein assay kits (Pierce).

*Statistics*
R and *P – values* for linear correlations were determined by ANOVA (analysis of variation) test (44)

*Acknowledgements*
We are grateful to Roy Kishony for great help at the early stages of this project, to Dan Tawfik, Art Horwich, Bill Eaton, Maxim Frank-Kamenetskii, Sergey Maslov and Stephen M. Gould for comments on the manuscript. We are grateful to Xiaowei Zhuang for help with FRET measurements. We thank Adrian Serohijos for the bioinformatics analysis, and Yakov

Pechersky, Abhishek Chintapalli, and Phil Snyder for invaluable technical assistance. This work was supported by NIH Grant No GM 068670 and long-term postdoctoral fellowship from the Human Frontier Science Program (SB).